# A taxonomic Approach to Topology Control in Ad-hoc and Wireless Networks

Matthias R. Brust and Steffen Rothkugel
*Faculty of Science, Technology and Communication (FSTC)*
*University of Luxembourg*
*Luxembourg*
*{matthias.brust, steffen.rothkugel}@uni.lu*

## Abstract

*Topology Control (TC) aims at tuning the topology of highly dynamic networks to provide better control over network resources and to increase the efficiency of communication. Recently, many TC protocols have been proposed. The protocols are designed for preserving connectivity, minimizing energy consumption, maximizing the overall network coverage or network capacity. Each TC protocol makes different assumptions about the network topology, environment detection resources, and control capacities. This circumstance makes it extremely difficult to comprehend the role and purpose of each protocol. To tackle this situation, a taxonomy for TC protocols is presented throughout this paper. Additionally, some TC protocols are classified based upon this taxonomy.*

## 1. Introduction

Multi-hop ad-hoc networks as well as wireless sensor networks are composed of a set of devices that communicate with each other over a wireless medium. Such networks can be formed spontaneously whenever devices are in transmission range. Joining and leaving of nodes occurs dynamically, particularly in the presence of mobility. Two mobile devices out of communication range can use intermediary devices for relaying packets. Such infrastructureless networks are very flexible and easy to deploy in different settings [1]. Potential applications of ad-hoc networks can be found in traffic scenarios, ubiquitous Internet access, collaborative work, and many more.

Those networks, however, suffer from unpredictable factors such as battery lifetime, interference, noise and—in particular—the dynamics in terms of potential mobile nodes or temporary link failures. An inappropriate topology can reduce the impact of network capacity by limiting spatial reuse of the communication channel and decrease network robustness.

Nevertheless, Topology Control picks up controllable factors such as the transmission range to tune the topology in order to get a more efficient communication network while inducing minimal additional overhead [2].

Recently, several TC protocols have been proposed. Designed for preserving connectivity, minimizing energy consumption, maximizing the overall network coverage or network capacity, each TC protocol makes different assumptions about the network topology, environment detection resources, and control capacities using different approaches, different models and having very different optimization objectives. This circumstance makes it extremely difficult to comprehend the role and purpose of each protocol. Due to the high number of characteristics, it has become a challenge to compare the performance of two or more protocols on a reasonable basis.

To overcome this difficult situation, we introduce a taxonomy for TC protocols. A taxonomy is a system for naming and organizing things (objects) into groups that share similar characteristics. A taxonomy helps organize knowledge that can be accessed, navigated, searched, discovered, and compared easier.

*Navigating.* Traversing the hierarchy of TC protocols; getting an overview of existing approaches in terms of quantity as well as quality.

*Searching.* Looking up for specific characteristics; applying querying mechanisms on the hierarchy.

*Comparing.* Elements that fit in one taxonomy are clearly comparable on the basis of the criteria of that taxonomy. Advantages and drawbacks of different approaches become visible by their lateral relationship Hence, a taxonomy is not a theoretical framework, but a method to make knowledge appropriately applicable.

*Discovering*. By spontaneously browsing through TC protocols something unexpected might be discovered. Deficiencies and shortcomings of TC protocols can be detected. Often new approaches are discovered by combining the categories of a taxonomy.

The contribution of this paper is to introduce a taxonomy for TC protocols focusing on three different purposes. The used categories or taxa are justified in a case study of TC protocols. As a proof-of-concept, a selection of the most important TC protocols is classified based upon this taxonomy.

The remainder of this paper is organized as follows. Section 2 describes taxonomic approaches of TC protocols and shows differences to our approach. Section 3 presents a case study of three TC protocols, namely LINT, LMST, and CBTC. In Sectio 4 the taxonomy is introduced, taxa are justified and described. This taxonomy is applied to the most important TC protocols in Section 5. This paper finishes with the conclusion and future work.

## 2. Related work

In this section, we describe existing taxonomic concepts and taxonomies of TC. We illustrate the differences between those approaches and ours.

One of the most detailed and applied taxonomy in the literature is introduced by Santi [3]. Santi's taxonomic approach is incremental. That means it is built on the previous state of the art of the TC techniques classifying the most important TC techniques. Although not explicitly given, the taxa appear to be controlling (action), quality of information (view), and computing character (decentralization). On the first level, there is a distinction between homogeneous and non-homogeneous TC. In the homogeneous case, the objective is all nodes adjust the transmission power to a minimum value where the network still remains connected. In the heterogeneous case, each node can adjust its transmission power to its own criteria. In contrast, our approach understands this distinction as part of the optimization criteria.

On the second level of this hierarchy, Santi categorizes TC techniques according to the quality of information necessary to run the protocol. With quality of information is meant the understanding that for instance location information or positions require different costs in terms of technology or communication than directional information of oncoming neighbors or just discovering neighbors by beaconing. Since Santi's taxonomy is incremental, it makes no difference in the quality of information between discovering neighbors and their distances. The reason is that technically by arranging the transmission range or by analyzing the signal strength of the received beacon message an estimation of the distance is feasible. Thus, techniques that require neighbor information or distances are categorized to the same class in Santi's taxonomy.

In contrast to Santi's taxonomy our approach was not incremental or hierarchical, but of generic and non-hierarchical nature. Please observe that both approaches are on a par, but necessarily leading to different frameworks. While Santi explicitly excludes clustering mechanisms, our approach aims to offer an interface for clustering mechanisms.

As a result of the generic approach, we differentiate between neighbor information and information of their distances. That is in contrast to [3]. Our assumption here is that distance information is of different quality than information where just the existence of certain neighbors is declared. The nature of the information is different.

In [3] a further distinction is considered. Hereby the type of communication/communication model is measured: *per-packet* and *periodic* topology control. Per-packet communication means that the transmission range of a node is adjusted according to the distance of the destination node to efficiently send a packet. In wireless medium however, neighbors are also able to receive that packet. The question arises why not use a periodic broadcast? This is called periodic topology control. The taxonomy presented in this paper does not consider a distinction between communication types, principally because per-packet communication just makes sense if you know the distance of the destination node. Thus, we consider the quality of information of higher importance.

Srivastava et al [1] presents a two-level hierarchical categorization system. The first level distinguishes between centralized and distributed algorithms and the second level focuses on evaluation and optimization criteria. Two evaluation criteria are explicitly introduced in the taxonomy: *connectivity awareness* and *capacity awareness*. Centralized approaches are not further classified, e.g. according to evaluation criteria. The stringent separation between connectivity and capacity awareness algorithms does not fully reflect the evaluation criteria of the LINT protocol [4] that focuses on minimizing energy consumption in a connected networks.

Contrary to [1] the taxonomy introduced in this paper avoids distinguish between centralization and distribution, because every centralized algorithm can be rewritten in its distributed equivalent [5]. We rather focus on a distinction between distribution and localization of an algorithm. Furthermore our taxonomy also introduces the achievement (i.e. guarantees and properties) of a protocol as own category. This is missing in Srivastava's et al [1].

## 3. Case study of TC protocols

This section describes three important and substantially different TC protocols in an informal but complete way. By fully describing the protocols we later extract important aspects that differentiate these protocols. These aspects are transformed into categories or taxa for developing the taxonomy in Section 4.

*LINT* [4]. The LINT (Local Information, No Topology) protocol aims to minimize average energy consumption, but at the same time preserving connectivity. It is explicitly designed for mobile ad-hoc networks. The approach has a heuristic nature that uses local information only. The basic idea of the protocol is that every node tries to keep the number of its neighbors close to an *ideal* number. Avoiding readjusting the transmission range for every change in the number of neighbors, readjustment is performed only if the number of neighbors is lower or higher than a minimum and maximum threshold. If the current number of neighbors is lower than the minimum threshold, the transmitting power is decreased. If the current number of neighbors is higher than the maximum threshold, the transmitting power is increased until the number of neighbors is lower than the high threshold or the maximum transmission range is reached. Decreasing and increasing happens in reasonable steps to minimize overhead. The message complexity depends on frequency of readjustment and is $n$ (number of nodes) for each readjustment of the whole topology. In fact, there is no recommendation for choosing the thresholds although they are critical parameters. Simulation results have been done using the random direction model. The results have shown that frequent changes in transmission power increases routing overhead and decreases throughput. Furthermore, for all cases there is an optimal frequency of topology checks where the throughput increases.

*LMST* [6]. The critical transmission range (CTR) problem describes the question of how to find a minimum transmission range assignment where each device in a static network is still connected. This problem can be optimally tackled by calculating the longest edge of the Euclidean minimum spanning tree (EMST) built on the nodes. As a result each node is assigned the same transmission range. The solution is optimal, but requires information of the entire network topology, i.e. global knowledge as well as node positions. This makes the EMST approach impractical. The LMST (Local Minimum Spanning Tree) protocol aims to approximate a minimum spanning tree with local knowledge only. The protocol consists of three phases: information exchange, topology construction, and determination of transmission power. In the first phase the beacon message is sent at maximum sending power. In phase two, each node receives the beacon message and constructs its local minimum spanning tree using Prim's algorithm. Thereby the links are weighted by their Euclidean distances. A new neighbor list is created. Here, a node $v$ is neighbor of node $u$ only if $v$ is a one-hop neighbor of $u$ in its minimum spanning tree. By measuring the signal strength of the beacons from phase one the sending power to reach any neighbor is determined. Finally, the sending power is set to the minimum to reach the farthest node in the new neighbor list. Observe that the resulting topology $G_{LMST}$ can be a directed graph. Applying one of the methods mentioned above/below results in an undirected topology $G_{LMST}+$ (turn unidirectional links into directional links) or $G_{LMST}-$ (removing unidirectional links). The LMST protocol can be computed in a fully distributed and localized manner and it preserves connectivity in the worst case. Computing $G_{LMST}$ requires sending only $n$ (number of nodes) messages.

*CBTC* [7]. The Cone-Based Topology Control (CBTC) protocol is designed for static network topologies. It works with any node deployment model. The optimization criteria are connectivity as well as minimizing energy consumption by removing energy inefficient links. The CBTC protocol divides the sensed environment in cones. The basic idea is to increase the transmission range up from an initial low value (ideally zero), but having at least one neighbor in each cone. Technically speaking directed antennas are able to cover cones. In CBTC each node can send a message in each direction. For dealing with the boundary problem, a so-called shrink-back operation is applied. The resulting topology $G_{CBTC}$ potentially contains asymmetric links. There are two classical ways to deal with asymmetric links: (a) augmentation of asymmetric links to symmetric links (*AugmCBTC*) and (b) removing asymmetric links (*RemCBTC*). Thus, for guaranteeing connectivity the cone size chosen is crucial. *AugmCBTC* has been proven to preserve network connectivity if $\alpha \leq 5\cdot\pi/6$. *RemCBTC* has been proved to preserve network connectivity if $\alpha \leq 2\cdot\pi/3$. The CBTC protocol is fully distributed which is based on local directional information only (i.e. localized), and preserves network connectivity for the optimal choice of the cone size. A weak point of CBTC is the high number of messages that must be exchanged to compute the network topology. First, there are message exchanges to figure out neighbors' directions. Second, adjustment of the transmit range to detect more neighbors causes additional message exchanges as well as the message exchange needed to render the symmetric topology.

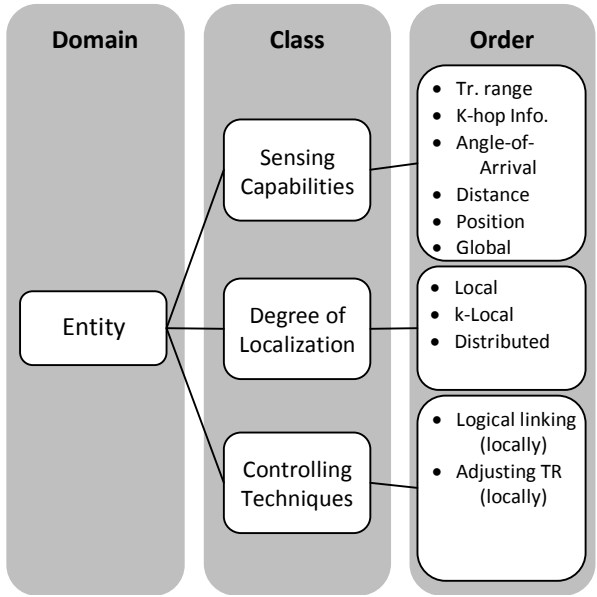

**Figure 1. The entity domain**

## 4. Taxonomy

A taxonomy cannot only be expressed as a hierarchical list of categories or taxa. Michel Foucault [8] argues that *things* can be ordered in any way. Concluding, there is no right way how to order *things*. For classification one has to emphasize certain characteristics and neglect other characteristics of a *thing*. But the overall objective is to order the *things* in a way that they appear more useful after classification. Additionally, they shall appear more useful when using the applied taxonomy instead of a different one.

The proposed taxonomy is of nonhierarchical functional order based on cybernetic ideas. That means that "records are categorized based on the functions and activities that produce them" [9]. Furthermore due to the complexity of TC protocols, the proposed taxonomy consists of three complementary areas or domains under investigation namely *entity*, *environment*, and *context*. Each domain takes account of a different "closed" view to TC protocols. The taxa used are called *domain*, *class*, and *order* (in hierarchical appearance).

*Domain*. A domain opens an almost complete view of an element, circumstance, or process. We describe the elements as *entities*, circumstances are described by the *environmental* domain, and the processes by the *context* domain. The taxon domain is answering the question: What is the area under investigation?

*Class*. A domain consists of classes. The classes are describing fundamental functions of TC protocols on a higher level in one domain. The question here is: What are the features, high-level attributes, etc. that all TC protocols have in common?

*Order*. The function of order can be interpreted as a possible state assignment of a class. Since we are dealing with a nonhierarchical taxonomy, multiple assignments are permitted. As we see below, multiple assignments are inevitable. Question: What are the individual class assignments of a certain TC protocol?

### 4.1. Entity

The domain entity describes the character of the considered element in the network. Looking from the graph theoretic point of view, an entity can be a node. A network designer considers a device as its entity of interest. Figure 1 shows the entity domain, its classes and orders. The appearance of the classes *Sensing Capabilities*, *Degree of Localization*, and *Controlling Techniques* is the result of a cybernetic view of TC.

**4.1.1. Sensing capabilities.** Each entity needs at least partial information of its environment in order to compute its action. For this, entities can access its sensing capabilities using the radio antenna to discover neighbors and querying neighbors for environmental data. We differentiate between primary and secondary sensing capabilities. Directly sensing the environment is done by primary sensing capabilities. Secondary sensing capabilities are environmental (e.g. topological) data provided by neighbors. By the use of secondary sensing capabilities, a device can augment its knowledge of the state of its environment. Primary sensing capabilities make use of technological interfaces. Wireless communication adapters, e.g. WiFi and Bluetooth, are used to discover neighbors within the transmission range and directed antennas make it feasible to discover the angle-of-arrival of a neighboring device [10]. GPS is used for position determination. These are just a few possibilities.

Typically, network equipment is more expensive than computing power. That is the reason why some approaches exist that substitute use of technological adapters with secondary sensing capabilities. For example, multi-lateration techniques create a relative positioning system by avoiding or minimizing the use of GPS adaptors in the network [11, 12]. The drawbacks are less position accuracy and additional use of communication and computing resources.

The reason why we create such a complex view of sensing is that environmental information provided to an entity can be interpreted as a kind of sensing information. Informally, it is an input for the protocol. Getting this information using primary or secondary sensing capabilities is of lower importance since often one can be substituted by the other. For example, the LMST protocol needs position information. There are two ways to get this information. An entity can request neighbors to send their GPS coordinates (use of secondary sensing capabilities, since environmental information is transferred). Second, the same entity can estimate the relative position of its neighbors by directed antennas and modifying transmission range (primary sensing capabilities). Neighbor discovery requires at least a beacon message from the neighbors and questions may arise if this process is a primary or secondary sensing? Beacon messages per se are just used for the neighbor discovering process, hence, are part of a primary sensing resource. Getting GPS data from the neighbor is a secondary sensing resource, because the requesting device is using the wireless communication channel to obtain GPS data. An entity getting neighbor lists from its neighborhood might discover the 2-hop environment even if the physical sensing capacities are restricted on the 1-hop neighborhood. The R&M protocol [13] shows that first the neighbors are sensed. The second phase of the protocol requires a global information exchange to create the optimal topology.

This class asks the questions: What kind of information does the TC protocol need (e.g. orders direction, speed, etc.) and how this information is gathered (e.g. by primary or secondary sensing capabilities)?

**4.1.2. Degree of localization.** For a TC protocol it is essential to be applicable in real world scenarios. In general we assume that the more a TC protocol limits its communication and computation demands the more it is applicable. From the nature of our area under investigation, the TC protocol must describe at least a distributed algorithm. Note that global algorithms are normally used to find the optimal solution (cf. CTR-problem, CNN-problem) in order to compare it to their distributed counterparts. Since every global algorithm can be rewritten into a distributed version [5], the concept of distribution appears not to be appropriate to our subject. Wattenhofer [5] suggests focusing on the communication of an entity and uses the model of localized and local algorithms as special cases of distributed algorithms. We do not differentiate between local and localized algorithms and the understanding of a *k*-localized algorithm as an algorithm where entities only communicate with their *k*-neighborhood (cf. [14]). Later, we identify the difference between local and localized algorithms as a fundamental part of the update policy, hence, not part of this class. Note that some approaches describe an algorithm as *fully* distributed when an algorithm is localized.

The class *Degree of Localization* identifies the (local) entity-based computing characteristics of the considered TC protocol. For example, LINT is a local protocol, because it uses the primary sensing capabilities, but avoids communication with the neighborhood.

Observe that in some approaches the information exchange related here to secondary sensing capabilities is seen as part of the computation. We strictly differentiate here. The degree of localization represents the exchange of computational information, e.g. exchanging a locally constructed topology like LMST. Clustering mechanisms create clusters with a clusterhead that often are 1-localized (e.g. [2]).

**4.1.3. Controlling techniques.** One of the most interesting aspects of a TC protocol is the entity's capacity to modify the topology in order to gain a desired global characteristic (cf. optimization criteria). Unfortunately, the ability to control the topology as a result of the execution of the TC protocol is extremely limited. In [4, 7] the transmission range is simply adjusted according to some criteria (magic number, CBTC one node in each direction, etc.). A more energy efficient approach is to maintain a fixed transmission range and remove undesired nodes from the neighbor list. This approach creates a logical topology with logical neighbors [15]. Clustering mechanisms are used in this way.

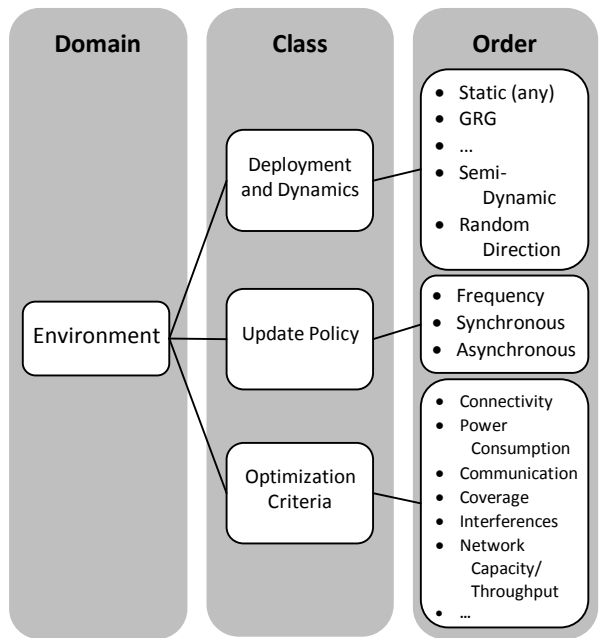

**Figure 2. The environment domain**

The class *Controlling Techniques* asks the question: What method or technique is used to construct the resulting topology?

## 4.2. Environment

The environment domain categorizes environmental characteristics of TC protocols. Essential differences like a static or mobile environment, coordination and frequency of execution phases as well as the optimization criteria are classes of the environment domain. In general this domain deals with global characteristics of participant entities. Figure 2 shows the environment domain, its classes and orders.

**4.2.1. Deployment and dynamic.** Wireless networks are an issue of dynamics. Mobile ad-hoc networks have to be investigated in the presence of node mobility. In contrast, wireless sensor networks where nodes might be static operate in a typically dynamic environment, because they are temporarily enabling/disabling links. Thus, dynamics is inherent to all forms of wireless networks. The dynamics of nodes directly impacts the performance of the protocols.

For simulation and analyses of mobile ad-hoc networks mobility models have been developed. Mobility models are a descriptive or probabilistic specification of possible or—ideally—realistic movement patterns, e.g. a pedestrian on a marketplace. One of the most used mobility models is the random waypoint (RWP) mobility model [16]. Entities are initially distributed uniformly at random in a unit square. Each entity chooses an arbitrary point at the unit square, a velocity between the parameters *min* and *max* as well as a pause time. After reaching its destination the entity waits and chooses the next destination. It is known that the RWP model has some drawbacks. The spatial node distribution generated by RWP is not uniform, but concentrated in the center of the unit square. Furthermore, the RWP model converges to that distribution just after a certain warming-up phase. The random direction model avoids the non-uniform spatial node distribution by just allowing the border as a potential destination for the entities. Someone may ask how realistic these mobility patterns are? Therefore, a variety of other mobility models have been developed, for instance Brownian-like motion, map-based, and group-based mobility models. For more information, we refer to [17, 18].

In the case of stationary networks, where entities are not able to move, the node deployment is the environmental condition that influences theoretical and simulation results. The node deployment is given by the graph model and used to describe the network links. For example, in geometric random graphs with an adapted model of random graphs, the entities are distributed according to some probability distribution in a certain region [19]. Other applied graph models are continuum percolation, geometric random graphs, and occupancy theory [15, 19, 20].

Above we affirmed that a TC protocol potentially assigns multiple orders. For single order assignment, node deployment and mobility models have to be separated. In fact, there are mobility models or which asymptotic distribution depends strongly on the initial node deployment. For instance, consider a mobility model in which nodes can move only within a circle of radius *r* around their initial position. The asymptotic distribution of this mobility model clearly depends on the initial deployment of the nodes, implying that the asymptotic distribution of the same mobility model with initial deployment is different from the asymptotic distribution with a different initial deployment.

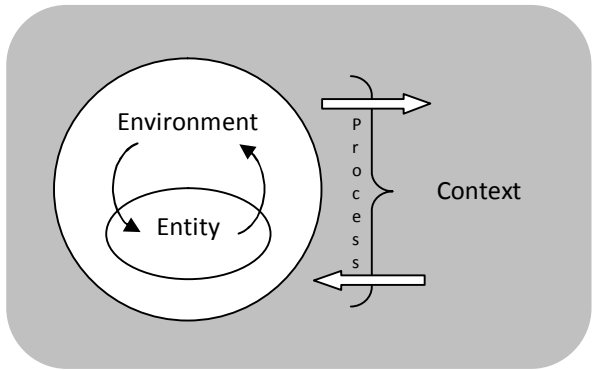

**Figure 3. The correlation between entity, environment and context**

Besides the general case, though in most cases of practical relevance (e.g., the well-known RWP model), the asymptotic node spatial distribution does not depend on the initial node deployment.

The class *Deployment and Dynamics* describes the spatial and temporal relationship of entities in an environment. Model driven parameters as density of nodes, velocity, pause time, obstacle-free simulation area, etc., are not part of our taxonomy but inherent to the model used. The question may arise why not include these parameters as taxa? The reason is that the characteristics of the protocol shall be of a more general nature. The mobility model or the node deployment, indeed, strongly influences the results because of the inherent probability property. Furthermore, variation in size, density, or velocity appears to be parameters of the mobility model, hence, have not been separated into its own taxon.

**4.2.2. Update policy.** The determination of the optimal frequency for reconfigurations is a hard-to-tackle problem [15]. In particular, when dealing with mobility, a delayed re-execution of the TC protocol results in an inappropriate topology. On the other side, a high frequency creates a control message overload as well as increases the power consumption with insignificant increases in the performance. The same problem appears for most clustering techniques.

The reconfiguration or re-execution phases of a TC protocol can be triggered synchronously as well as asynchronously. Asynchronous execution means that each entity decides by itself when to start a new phase. This decision might potentially also include waiting for a notification from a neighboring node, thus including dependencies. In the synchronous case, we assume the following restriction. An entity's operations in phase $i$ may only depend on the information received during phases 1 to $i-1$. In fact this is the difference between localized and local algorithms introduced but [5] ignored in the class degree of localization. Although we understand it as part of the update policy.

Synchronous phases are making more requirements on the media access layer. It is costly to implement a media access control scheme that allows synchronous topology reconfigurations. Messages are getting lost due to interferences or dynamics. Asynchronous execution shows potentially different results than its synchronous equivalent. Watts [14] shows this difference by executing genetic algorithms in a small-world topology by using synchronous phases as well as asynchronous ones.

The class *Update Policy* asks about what the character of the entities state transition from an environmental point.

**4.2.3. Optimization criteria.** Each TC protocol is designed based on an optimization criterion. In order to guarantee message delivery a fundamental issue is to provide connectivity in a network, i.e. each entity is reachable. The connectivity criterion can be more restrictive when requiring at least two distinct routes from source to destination. This demand for a 2-connectivity (in general terms $k$-connectivity) increases the reliability of a network in respect to message delivery and fault tolerance. The connectivity criterion does not make sense if the entities have just transmission range as sensing capabilities, because the optimal solution is to send with maximum transmission range. Another example is minimizing energy consumption by reducing the sending range to zero.

Thus, TC protocols are designed following multiple optimization criteria. That is an additional reason why it needs multiple order assignments. As seen, LINT aims to minimize average energy consumption, but at the same time keep connectivity. In contrast to wired point-to-point channels, wireless communications use the shared radio channel. A shared communication medium implies that care must be paid to prevent the concurrent wireless transmissions from corrupting each other. These interferences decrease the network capacity drastically and lead to some protocols using it as an additional optimizing criterion (see also LINT and LMST). Other protocols pay attention to optimizing the communication by analyzing the communication pattern and data flow [13] or maximizing the coverage area dealing namely with the Critical Coverage Range (CCR) problem.

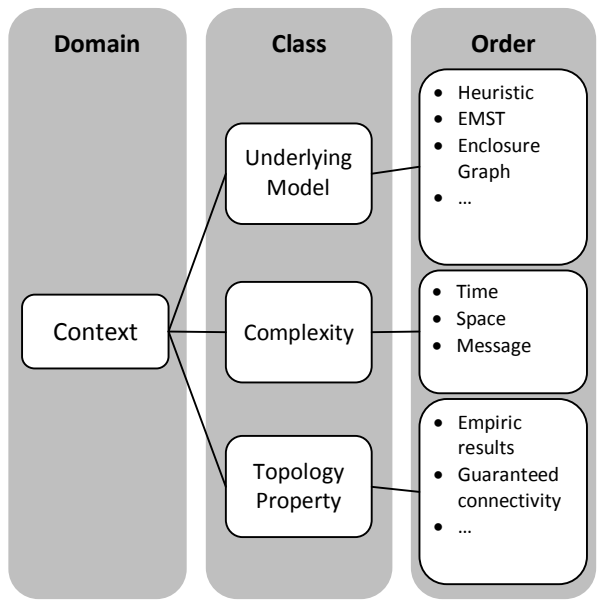

**Figure 4. The context domain**

The variety of possible combinations of optimization criteria almost impedes the comparison of TC protocols. Therefore, the class *Optimization Criteria* put in question the criteria that a TC protocol focuses on.

### 4.3. Context

The entity interacts with its environment causing a process that we call context. The context is of holistic character. The context is a process that is mainly based on the underlying model that is used to design the TC protocol as well as the outcomes of that process. The outcomes are the complexity as well as the topology properties of the TC protocol. The correlation between entity, environment, and context is illustrated in Figure 3 the context domain, its classes, and order are illustrated in Figure 4.

**4.3.1. Underlying model.** When designing a TC protocol the design follows a model or hypothesis or at least is inspired by some similar assumptions. This procedure influences the internal structure of the protocol and its performance. For example the LMST protocol approximates the optimal solution known as Euclidean Minimum Spanning Tree. The class *Underlying Model* asks which model is approach based. This model can also be heuristic, using intuition for justification of the approach. As the name implies LINT is a heuristic approach using no model for justification of the approach.

**4.3.2. Complexity.** As already mentioned, the performance is an important category when we are talking about TC protocols. The performance can be divided into the complexity properties and topology properties. Since we are operating in a network, besides the common time and space complexity, the message complexity appears as an important property of a protocol. For instance, the message complexity of LINT is *n* (number of nodes).

The class *Complexity* asks the question about how much communication and computation effort in terms of the O-notion is necessary to run that TC protocol.

**4.3.3. Topology property.** The CBTC protocol is of heuristic nature, but there exists some proof in regard to connectivity and behavior of the protocol. These properties of the resulting topologies are a performance measure. Therefore the protocol properties are a fundamental class in our taxonomy. The stretch factor measurement is used to compare the quality of the constructed topology respective of some criteria. Thus, there is a distance stretch factor and a hop stretch factor. The power stretch factor measures the ratio of the *maxpower* graph (when all nodes send with maximum transmission power) and the power graph of the resulting topology. With these measurements it becomes feasible in terms of quantity to see which characteristics of a topology are becoming more efficient and which characteristics suffer from this improvement.

The class *Topology Property* describes the quality of the expected topology constructed by a certain TC protocol.

### 4.4. Discussion

**4.4.1. Asymmetric links.** Some protocols as CBTC [7] produce directed graphs. The question arises as to why this characteristic of the resulting topology isn't considered as criterion for the taxonomy. If necessary to produce a bidirectional graph, asymmetric links are often removed or augmented as part of the proper protocol. Thus, the problem can be tackled with additional communication and computation effort and, hence, is of lower relevance for being a taxon.

## Table 1. Classification of TC protocols

| Domain | Class | Order | R&M | LINT | CLTC-A | LMST | K-Neigh | CBTC |
|---|---|---|---|---|---|---|---|---|
| Entity | Sensing Capabilities | Transmission range | X | X | | X | X | |
| | | Angle-of-Arrival | | | | | | X (cones) |
| | | Position | | | X | | | |
| | | Global | X | | | | | |
| | | Distance | | | | X (implicit) | X (estimation) | |
| | Degree of Localization | Local | | X | | X | X | X |
| | | k-Local | | | X | | | |
| | | Distributed | X | | | | | |
| | Controlling Techniques | Logical Linking | | | | X | | X |
| | | Adjusting TR | X | X (k-neighbors) | X | | X | X |
| Environment | Deployment and Dynamics | Static (GRG,…) | | | X | | | |
| | | Static (Any Node-deployment) | X | | | X | X | X |
| | | Mobile (Random Direction Model) | | X | | | | |
| | Update Policy | Synchronous | X | | | X | X | |
| | | Asynchronous | | | | | | |
| | | Frequency | | | | | | |
| | Optimization Criteria | Energy consumption | X | X | X | X (simulative) | X | X |
| | | Connectivity | | X | X (k-conn) | X (1-conn) | | X |
| | | Efficient communication | X (all-to-one pattern) | | | | | |
| | | Fault tolerance | | | | X | | |
| | | Network capacity/Throughput | | X | | X (simulative) | | |
| Context | Complexity | Time-complexity | | | | Prime-Alg. | | |
| | | Space-complexity | | | | Prime-Alg. | | |
| | | Message-complexity | | n (number of nodes) | | | 2n | |
| | Underlying Model | Enclosure Graph | X | | | | | |
| | | Heuristic | | X | | | | |
| | | CTR-Problem; EMST | | | | X | | |
| | | $G_k^-$ | | | | | | X |
| | Topology Property | | Strong connectivity Optimal topology for MinEnergyAllToOne problem | Empirical results: optimal frequency for topology checks | k-connectivity Scalability | Worst-case connectivity Assym: GLMST; Sym: GLMST+;GLMST- | Does not preserve connectivity in the worst case 20% less energy use than CTBC | Preseve worst-case connectivity basicCBTC augmCBTC remCBTC |

**4.4.2. Topology Control vs. Clustering Mechanism.** Some authors make a strong separation between topology control and clustering mechanisms mainly because clustering mechanisms provide a hierarchical arrangement of nodes. For example, the clusterhead is a central communication point for slaves, etc. In its essence, clustering mechanisms construct a logical topology like CBTC. Although TC protocols do not deal with hierarchical node arrangement, this paper considers clustering a particular part of TC. Thus, the taxonomy was also designed in order to include clustering mechanisms. Further testing will prove out the taxonomy in terms of clustering.

## 5. Applying the taxonomy

The produced taxonomy has been applied with success to a variety of different TC protocols. Table 1 illustrates the LMST, LINT, and CBTC protocols as described in Section 3. Additionally, the K-Neigh, R&M, and CLTC-A protocols are classified. Note that the protocols in Table 1 are described exemplary, but not complete.

A widely studied TC protocol is the K-Neigh protocol introduced in [21]. This protocol maintains the number of physical neighbors equal to (or slightly below) $k$. The number $k$ is chosen in such a way that the graph generated is connected with high probability. This protocol produces symmetric topology and is more energy-efficient than CBTC but it does not ensure the connectivity all the time due to probability-based algorithm.

The R&M protocol [13] uses the notion of relay region and enclosure for the purpose of energy control. Instead of transmitting directly, a node chooses to relay through other nodes in case that less energy will be consumed. It is shown that the network is strongly connected if every node maintains links with the nodes in its enclosure and the resulting topology is a minimum power topology.

The Cluster-based Topology Control (CLTC) framework [2] is a k-localized approach using transmission power adjustment. CLTC employs a clustering algorithm and each cluster locally exchanges information among its own members and with neighboring clusters to achieve *k*-connectivity. It is shown that the CLTC framework guarantees global *k*-connectivity. Furthermore, the framework allows TC algorithms with different optimization criteria.

## 6. Conclusion and future work

This paper presented a taxonomy for Topology Control (TC) protocols. The complexity of this taxonomy represents the innumerous distinctive characteristics of TC protocols. We defined taxa, namely entity, environment and context and explained the significance in the field of TC. Additionally as proof-of-concept we classified the most important TC protocols in that taxonomy for validation reasons. We believe that this work helps us to understand each aspect of TC protocol design better and assists in organizing the high number of TC protocols.

## 7. References

[1] G. Srivastava, P. Boustead, and J. F. Chicharo, "A Comparision of Topology Control Algorithms for Ad-hoc Networks," in *2003 Australian Telecommunications, Networks and Applications Conference*, 2003.

[2] C.-C. Shen, C. Srisathapornphat, R. Liu, Z. Huang, and E. L. Lloyd, "CLTC: A Cluster-Based Topology Control Framework for Ad Hoc Networks," *IEEE Transactions on Mobile Computing,* vol. 03, pp. 18-32, 2004.

[3] P. Santi, "Topology Control in Wireless Ad Hoc and Sensor Networks," *ACM Computing Surveys,* vol. 37, pp. 164-194, 2005.

[4] R. Ramanathan and R. Rosales-Hain, "Topology control of multihop wireless networks using transmit poweradjustment," in *19th Annual Joint Conference of the IEEE Computer and Communications Societies*, 2000.

[5] R. Wattenhofer, "Sensor Networks: Distributed Algorithms Reloaded - Or Revolutions?," in *13th Colloquium on Structural Information and Communication Complexity (SIROCCO)*, Chester, United Kingdom, 2006.

[6] N. Li, J. C. Hou, and L. Sha, "Design and analysis of an MST-based topology control algorithm," in *INFOCOM 2003. Twenty-Second Annual Joint Conference of the IEEE Computer and Communications Societies*, 2003.

[7] L. Li, J. Halpern, V. Bahl, Y.-M. Wang, and R. Wattenhofer, "Analysis of a Cone-Based Distributed Topology Control Algorithm for Wireless Multihop Networks," in *Twentieth ACM Symposium on Principles of Distributed Computing (PODC)*, 2001.

[8] M. Foucault, "The Order of Things: An Archaeology of the Human Sciences. 1966," 1973.

[9] B. Blackburn, "Taxonomy Design Types," *The ECM Association,* 2006.

[10] S. Basagni, S. Giordano, I. Stojmenovic, and M. Conti, *Mobile Ad Hoc Networking*: Wiley-IEEE Press, 2004.

[11] M. Hamdi, S. Capkun, and J. P. Hubaux, "GPS-free Positioning in Mobile Ad-Hoc Networks," in *Proceedings of HICSS*, Hawaii, 2001.

[12] D. Niculescu and B. Nath, "Ad hoc positioning system (APS)," in *INFOCOM* San Francisco, 2003.

[13] V. Rodoplu and T. H. Meng, "Minimum energy mobile wireless networks," *IEEE Journal on Selected Areas in Communications,* vol. 17, pp. 1333-1344, 1999.

[14] D. J. Watts, *Small Worlds*: Princeton Studies in Complexity, 1999.

[15] P. Santi, *Topology Control in Wireless Ad Hoc and Sensor Networks*: Wiley, 2005.

[16] J. Yoon, M. Liu, and B. Noble, "Random waypoint considered harmful." INFOCOM 2003. pp. 1312- 1321.

[17] T. Camp, J. Boleng, and V. Davies, "A survey of mobility models for ad hoc network research." vol. 2, 2002, pp. 483-502.

[18] V. A. Davies, "Evaluating mobility models within an ad hoc network," advisor: Tracy Camp, Dept. of Mathematical and Computer Sciences.Colorado School of Mines, 2000.

[19] M. Penrose, *Random Geometric Graphs*: Oxford University Press, 2003.

[20] P. Hall, "On Continuum Percolation," *The Annals of Probability,* vol. 13, pp. 1250-1266, 1985.

[21] D. Blough, M. Leoncini, G. Resta, and P. Santi, "The K-Neigh Protocol for Symmetric Topology Control in Ad Hoc Networks," in *ACM MobiHoc 03*, 2003, pp. 141-152.